\documentstyle[prd,aps,floats,epsf]{revtex}

\newcommand{\BH}	{{\rm BH}}
\newcommand{\planck}	{{\rm pl}}
\newcommand{\radiation}	{{\rm rad}}
\newcommand{\weak}	{{\rm W}}
\newcommand{\DW}	{{\rm DW}}
\newcommand{\TeV}	{{\rm \;TeV}}
\newcommand{\GeV}	{{\rm \;GeV}}
\newcommand{\MeV}	{{\rm \;MeV}}

\newcommand{\kg}	{{\rm \;kg}}

\newcommand{\univ}	{{\rm univ}}

\newcommand{\sph}	{{\rm sph}}
\newcommand{\lnear}{{\;\mbox{\raisebox{-1mm}{$\sim$}\hspace{-2.8mm}\raisebox{0.7mm}{$<$}}\;}}
\newcommand{\gnear}{{\;\mbox{\raisebox{-1mm}{$\sim$}\hspace{-2.6mm}\raisebox{0.7mm}{$>$}}\;}}

\newcommand{\fig}[1]	{Figure \ref{#1}}

\newcommand{\PRD}[3]	{{Phys. Rev. D}       {\bf #1}, #2 (#3)}
\newcommand{\PLB}[3]	{{Phys. Lett. B}      {\bf #1}, #2 (#3)}
\newcommand{\PL}[3]	{{Phys. Lett.}        {\bf #1}, #2 (#3)}
\newcommand{\NuP}[3]	{{Nucl. Phys.}        {\bf #1}, #2 (#3)}

\newcommand{\Nature}[3]	{{Phys. Rev. Lett.}   {\bf #1}, #2 (#3)}

\draft
\begin{document}
\twocolumn[\hsize\textwidth\columnwidth\hsize\csname
@twocolumnfalse\endcsname

\preprint{DPNU-98-19}
\title{Electroweak baryogenesis by black holes}
\author{Yukinori Nagatani \cite{nagatani}}
\address{Department of Physics, Nagoya University, Nagoya 464-8602, Japan}
\date{November 27, 1998}
\maketitle


\begin{abstract}
We propose a new scenario of the electroweak baryogenesis
based on the properties of electroweak domain walls surrounding black holes
in the Higgs phase vacuum of the extended Standard Model with 2-Higgs doublets.
It is shown
that there exists the electroweak domain walls and the symmetric regions
surrounding black holes
and that the Hawking radiation of the black holes can produce
a non-zero baryon number density.
The scenario can explain the origin of the baryon number in our universe
even without the assumption of the existence of the first order electroweak
phase transition.
\end{abstract}
%
%
\vskip1.9pc]
\narrowtext


The thermal radiation from a black hole was discovered
by Hawking \cite{Hawking},
and contributions of primordial black holes in the early universe
have been discussed \cite{Primordial}.
Then it is natural to ask if black holes can play a role in the baryogenesis.
Barrow et al. pointed out this possibility
based on the GUT baryogenesis \cite{Barrow},
but these theory cannot avoid
the washout problem by the sphaleron process \cite{Sph}.
To avoid this problem,
Majumdar et al. consider the GUT-baryogenesis by black holes
survived in the symmetric phase vacuum \cite{Majumdar}.
Because of the washout problem,
the electroweak baryogenesis proposed by Cohen et al. \cite{CKN}\cite{CKN2}
is an important scenario,
in which the electroweak domain wall
created by the assumed first order phase transition
plays a crucial role.

In this paper, we show two new and important points:
First,
we find existence of the electroweak domain wall and the symmetric region
surrounding the black hole by the Hawking radiation;
Second, by the previous mechanism,
we show the possibility of the {\it electroweak} baryogenesis
without the first order phase transition
in the Higgs phase vacuum of the extended Standard Model (SM)
with two Higgs doublets.
We then propose new scenario of the baryogenesis by the black holes,
which can create the baryon number
with the baryon-entropy ratio $B/S \simeq 10^{-9} \sim 10^{-10}$
in the early universe,
and can satisfy the requirement of the big-bang nucleosynthesis (BBN),
if most of the matter existed as primordial black holes
with mass of some hundreds kilograms
and our domain wall has a CP phase $\Delta\theta/\pi \gnear 0.1$.

The most important difference
between our scenario and the existing one \cite{CKN}
is the requirement on the nature of the phase transition:
In the scenario discussed in Ref.\cite{CKN},
one needs the first order phase transition to
create the domain wall to realize non-equilibrium,
and the structure of the domain wall is determined by the
dynamics of phase transition.
In our scenario, {\it we do not need the first order phase transition},
since the thermal structure of the black hole creates the domain wall
and determines its structure.
Finally, we point out that the existence of the symmetric region
surrounding the black hole washes out the baryon number
created by the GUT-baryogenesis discussed in Ref.\cite{Majumdar}.


First,
we discuss how to satisfy the Sakharov's three conditions \cite{Sakharov}
for the baryogenesis.
Let us consider the Schwarzschild black holes
whose Hawking temperature $T_\BH$ is higher than
the critical temperature of electroweak phase transition.
In the Higgs phase vacuum,
the radiation from these black holes restores the electroweak symmetry
in the neighborhood of the horizon
and the electroweak domain wall does appear
as we shall demonstrate below.
Then we can discuss the baryogenesis scenarios
in analogy with the ordinary electroweak baryogenesis \cite{CKN}.
Here
we assume the two-Higgs-doublets extension of the Standard Model (2HSM)
as the background field theory for the origin of CP phase in the domain walls,
and that the electroweak phase transition is the second order
with its critical temperature taken as $T_\weak = 100 \GeV$
for simplicity.
Actually,
the Sakharov's  three conditions \cite{Sakharov}
for baryogenesis are satisfied as follows:
\begin{enumerate}
 \item The baryon number violation: 
       The sphaleron process \cite{Sph} takes place
	in the domain wall near the symmetric region.
 \item The C-asymmetry: the SM is the chiral theory.\\
       The CP-asymmetry: Here we assume in 2HSM that
       the domain wall has the space-dependent CP phase.
 \item Out of equilibrium:
       The black-hole radiation is a non-equilibrium process.
       This radiation creates the spherical domain wall
       and the radiated particles pass through.
\end{enumerate}


To begin with we note that
the Schwarzschild black hole mass $m_\BH$, temperature $T_\BH$,
Schwarzschild radius $r_\BH$, and lifetime $\tau_\BH$
are related by the equations:
$
 T_\BH		=  \frac{1}{8 \pi} \frac{m_\planck^2}{m_\BH},\ 
 r_\BH		= 2 \frac{m_\BH}{m_\planck^2}
 		= \frac{1}{4 \pi} \frac{1}{T_\BH},\ 
 \tau_\BH	\simeq  \frac{10240}{g_*} \frac{m_\BH^3}{m_\planck^4}
		= \frac{20}{\pi^2 g_*} \frac{m_\planck^2}{T_\BH^3},
$
where $g_*$ is the freedom of the massless particles that
this black hole can decay into at its temperature.
In the electroweak critical temperature, 
we have $g_* \simeq 100$.
In this paper,
we parameterize black holes by its Hawking temperature
rather than its mass for convenience.
We display these relations in \fig{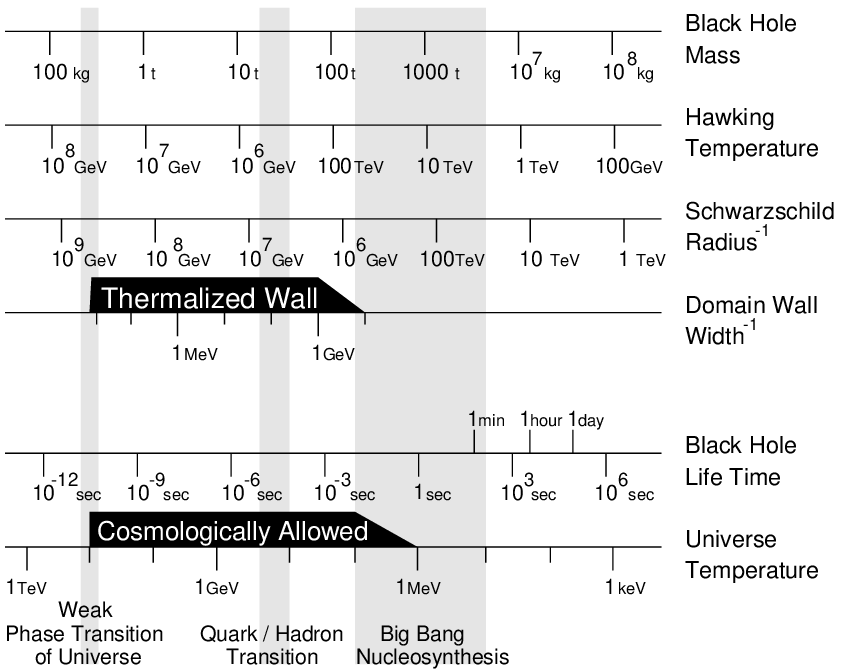}.

\begin{figure}
\begin{center}%
 \ \epsfbox{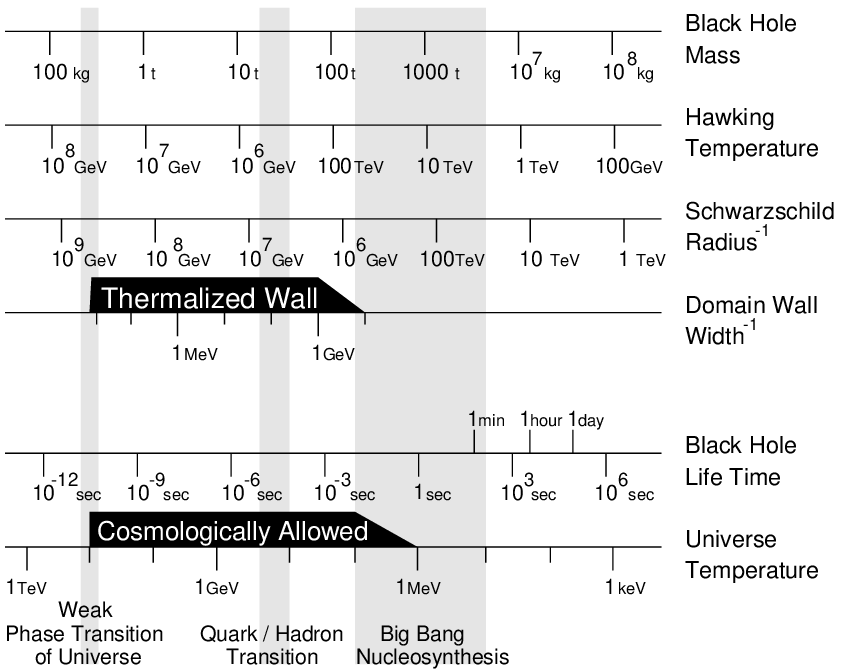}
 \caption{The relations between parameters of the Schwarzschild black hole.
 	  In the relation between the black-hole lifetime and
 	  the universe temperature,
 	  we used correspondence between
 	  the black-hole lifetime and the age of the universe,
 	  because we assumed that
 	  many primordial black holes had existed in the very early universe
 	  and evaporated at the age of the universe.
          }%
          \label{BHTemp.eps}%
\end{center}%
\end{figure}

Now,
we consider the space-dependence of the temperature
in the neighborhood of the black hole
by the local thermal equilibrium (LTE) approximation.
Because of the spherical symmetry of the black hole,
we put the local temperature as $T(r)$,
where $r$ means the radius from the center of the black hole.
In the symmetric phase,
the mean free path (MFP) of quarks and gluons
is determined by QCD interaction and
the MFP is a function of the temperature:
$\lambda_s(T) = \beta_s/T$, where $T$ is the temperature of background plasma
and $\beta_s \simeq 10$ is a constant related to the QCD coupling constant
$\alpha_s$. 
The Hawking radiation thermalizes, at first, a sphere
with the radius $r_s = \lambda_s(T_s)$ to the temperature $T_s$,
because the size of the black hole is smaller than the MFP:
$r_\BH < \lambda_s(T_\BH)$.
We call this radius $r_s$ a minimal thermalized radius.
Later we determine this radius $r_s$.
For every sphere greater than the minimal thermalized radius,
we can make sure of the thermalization of such a sphere
by the temperature distribution $T(r)$.

To determine the temperature distribution $T(r)$,
we discuss the transfer equation of energy.
The energy diffusion current in the LTE is
$J_\mu = -\lambda(T) \: \partial_\mu (\frac{\pi^2}{120}{g_*}T^4)$,
where $\lambda(T) = \beta/T$ is the effective MFP of all particles
by all interactions of the SM,
and we can estimate $\beta \simeq 100$.
Then the transfer equation is $\dot{\rho} = \nabla_\mu J^\mu$,
where $\rho = \frac{\pi^2}{30} {g_*}T^4$ is the energy density.
This treatment is referred to as the diffusion approximation of
photon transfer at the deep light-depth region \cite{Mihalas}.
The stable spherical-symmetric general solution of this equation is
$T^3(r) = T_\univ^3 + (T_s^3 - T_\univ^3) \frac{r_s}{r}$,
where $T_\univ = T(r \rightarrow \infty)$ is the background universe
temperature.
Here,
we approximate that the freedom of the massless particles ${g_*}$ is constant,
because we consider only $T(r) \gnear T_\weak$.
Our stable solution has total out-going energy flux $r$-independent:
$F = 4\pi r^2 \times J_r = \frac{2\pi^3}{45}\beta_s\beta c_s {g_*}T_s^2$,
where $c_s = 1 - (T_\univ/T_s)^3$ is a factor of the background correction.
This flux must be equal to the total flux by the Hawking radiation:
$F_\BH = 4\pi r_\BH^2 \times \frac{\pi^2}{120} {g_*}T_\BH^4$.
This relation $F = F_\BH$
leads us to the temperature of the minimum thermalized sphere:
$T_s = \frac{\sqrt{3}}{8\pi\sqrt{\beta_s\beta}\sqrt{c_s}} T_\BH$.
Finally we get the spherical thermal distribution
surrounding the black hole:
\begin{eqnarray}
T(r) &=& \left[
	  T_\univ^3 + \frac{3}{64\pi^2}\frac{1}{\beta}\frac{T_\BH^2}{r}
	 \right]^{1/3}
\end{eqnarray}
for $r > r_s = \frac{8\pi\sqrt{\beta_s^3\beta}\sqrt{c_s}}{\sqrt{3}}\frac{1}{T_\BH}$.


Let us next discuss formation of the domain wall.
The local temperature $T(r)$ is a decreasing function of $r$.
This local temperature configuration allows us
a nontrivial phase structure of the symmetry breaking depending on the space,
i.e., the vacuum expectation value (VEV) of Higgs doublets
depends on the distance from the center of the black hole $r$.
By the LTE
the space dependence of the Higgs doublets may take the form
\begin{eqnarray}
 \langle\phi_i(r)\rangle &=& \langle\phi_i\rangle_{T=T(r)}.
\end{eqnarray}
For simplicity,
we assume that the electroweak phase transition is the second order
and the simplest form of the Higgs VEV as
$
 (|\langle\phi\rangle_T|/v)^2 + (T/T_\weak)^2 = 1.
$
Then the CP-phased neutral Higgs VEV
may be written as
\begin{eqnarray}
 \langle\phi_1^0(r)\rangle &=&
  \left\{
   \begin{array}{lcl}
    0 & & (r \leq r_\DW) \\
    v_1 f(r) \; e^{-i \Delta\theta (1-f(r))} & & (r > r_\DW)
   \end{array}
  \right.,
\end{eqnarray}
where $f(r) = \sqrt{1 - (T(r)/T_\weak)^2}$ (see \fig{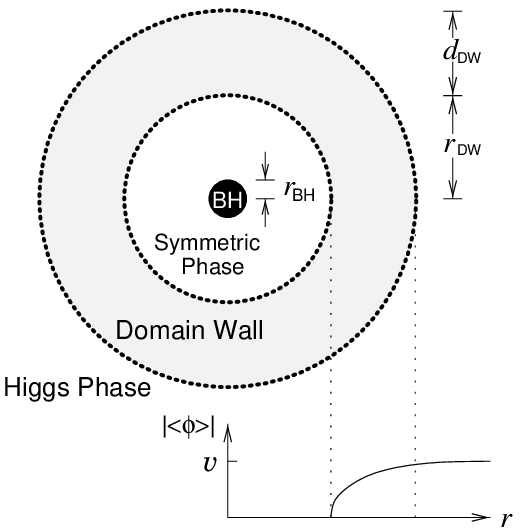}).
In this configuration of the Higgs VEV,
the width of our domain wall $d_\DW$
is approximately equal to the radius of the symmetric region $r_\DW$.
By $T(r_\DW) = T_{\weak}$, we find
\begin{eqnarray}
 d_\DW \simeq r_\DW
  &=& \frac{3}{64 \pi^2} \frac{1}{\beta c_\weak} \frac{T_\BH^2}{T_\weak^3},
\end{eqnarray}
where $c_\weak = 1 - (T_\univ/T_\weak)^3$ is other background correction.
We see that this electroweak phase structure is determined by
the thermal structure of the black hole.
We illustrate this width in \fig{BHTemp.eps}.

\begin{figure}
\begin{center}%
 \ \epsfbox{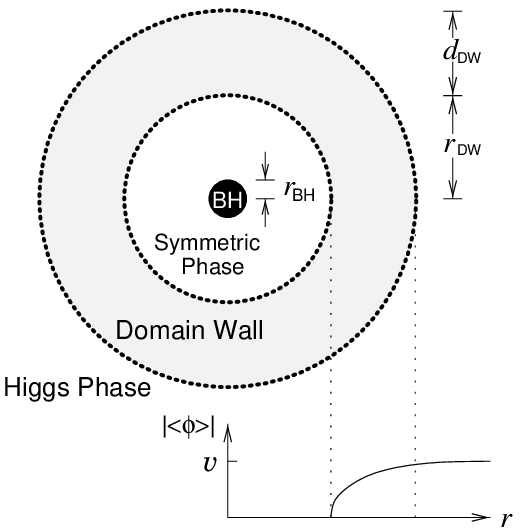}
 \caption{The electroweak phase structure depending on the space
	  surrounding the black hole.
 	  We note that $r_\BH \ll r_\DW \simeq d_\DW$.
          }%
          \label{BHWall.eps}%
\end{center}%
\end{figure}

On the other hand,
in the ordinary electroweak baryogenesis scenarios \cite{CKN},
structure of the domain wall
is determined by the dynamics of phase transition,
and they assumed $f(z) = \frac{1}{2}[\tanh{z/\delta} + 1]$,
where $z$ is the perpendicular direction to wall and
$\delta$ is the width of the wall.


The mean velocity of the out-going diffusing particles at radius $r$ is 
$v(r) \simeq J_r(r)/\rho(r)$.
Especially, the mean velocity at the domain wall is
$v_\DW = v(r_\DW) \simeq \frac{64 \pi^2}{9} \beta^2 c_\weak^2
 \left(\frac{T_\weak}{T_\BH}\right)^2$.
This means that
the plasma radiated from the black hole is
flowing from the symmetric region to the broken one
with velocity $v_\DW$ at the domain wall,
i.e., the plasma and domain wall moves relatively.

Our stable LTE approximation for baryogenesis is valid
when the size of the domain wall is greater than the MFP
and when the black-hole lifetime is large enough
to keep the stationary electroweak domain wall.
The first condition is
$1 \ll d_\DW/\lambda(T_\weak)
 = (\frac{\sqrt{3}}{8\pi\sqrt{\beta_s\beta}\sqrt{c_\weak}}
    \frac{T_\BH}{T_\weak})^2$,
and hence we have $T_\BH \gnear 4.6\times10^4\GeV$.
Because 
the time to construct the stable weak domain wall is
$\tau_\DW \simeq r_\DW / v_\DW
 = \frac{27}{4096 \pi^4}
   \frac{1}{\beta^3 c_\weak^3}
   \frac{T_\BH^4}{T_\weak^5}$,
the second condition leads us to
$1 \ll \tau_\BH / \tau_\weak
   \simeq \frac{81920 \pi^2}{27 {g_*}} \beta^3 c_\weak^3
          \frac{m_\planck^2 T_\weak^5}{T_\BH^7}$.
Then we get a restriction for the black-hole temperature as
$T_\BH \lnear 3.9 \times 10^7 \GeV$.
To obtain this restriction,
we used the later relation
between the Hawking temperature and the universe temperature.
We note that we have $T_\univ \simeq 98\GeV$,
when $T_\BH = 3.9 \times 10^7 \GeV$.
When these two conditions are satisfied,
we can easily check the thermalization of the domain wall
in the meaning of time scale.
We illustrate these conditions in \fig{BHTemp.eps}
as the thermalized wall.


Here
we propose a scenario of the baryogenesis due to these black holes.
Because the width of domain wall $d_\DW$
is greater than the MFP of the quarks $\lambda(T_\weak)$,
and particles at domain wall have a mean out-going velocity $v_\DW$,
then we can consider 
a variant of spontaneous baryogenesis scenario \cite{CKN},
which we call the ``thick-wall black-hole baryogenesis''.

The C- and CP-asymmetries take place in the domain wall
as the space-dependent physical CP phase of the domain wall
and the baryon-number-violating process also occurs in the domain wall near
the symmetric phase as the sphaleron one.
Because the domain wall is thermalized,
the sphaleron transition rate in the domain wall is
$
 \Gamma_\sph = \kappa \alpha_\weak^5 T_\weak^4 e^{-E_\sph/T_\weak},\ 
 E_\sph(T) = \frac{2M_W(T)B}{\alpha_\weak} \simeq
  \frac{|\langle\phi\rangle|_T}{v_1} \times 10 \TeV,
$
where $\kappa \sim O(30)$ is a numerical constant \cite{SphRate}.
When $|\langle\phi(r)\rangle|/v_1 < \epsilon \simeq 1/100$,
i.e., $E_\sph < T_\weak$,
the sphaleron transition is not suppressed by the exponential factor.
Then we can consider the sphaleron process
only in the neighborhood of the symmetric region in the domain wall.
We define the width of this region $d_\sph$
by $f(r_\DW+d_\sph) = \epsilon$ because of the form of Higgs VEV.
Then the volume integral of the sphaleron transition at work is
$ V = 4\pi r_\DW^2 \times \int_{r_\DW}^{r_\DW+d_\sph} dr$.
In this region, we have a space-dependent CP phase $\theta(r)$,
and diffusing particles have an out-going mean velocity $v_\DW$.
Then at the co-moving frame of this plasma-flow,
these particles feel the time-dependence of the CP phase:
$ \dot{\theta} \simeq v_\DW \frac{d}{dr} \theta $.
In the ordinary spontaneous baryogenesis scenario,
the plasma containing top quarks is at rest but the domain wall is moving,
while in our scenario,
the domain wall is rest but the plasma is flowing though the domain wall.
The relation between the baryon-number chemical potential
and the time-dependent CP phase is
$ \mu_B = {\cal N} \dot{\theta} $,
where ${\cal N} \sim O(1)$ is a model-dependent constant \cite{CKN}.
Finally, we can write down
by the detailed-balance relation
the rate of the baryon number creation per black hole:
\begin{eqnarray}
 \dot{B}
 &=& - V \; \frac{\Gamma_\sph}{T_\weak} \; \mu_B
       \nonumber\\
 &=& - 4\pi {\cal N} \kappa \alpha_\weak^5 T_\weak^3 \;
       r_\DW^2 \; v_\DW
       \int_{r_\DW}^{r_\DW+d_\sph} dr \: \frac{d}{dr} \theta(r)
       \nonumber\\
 &=& - \frac{1}{16\pi} \: {\cal N} \kappa \alpha_\weak^5 \:
       \epsilon\Delta\theta \:
       \frac{T_\BH^2}{T_\weak},
\end{eqnarray}
where we have used the relation
$\int_{r_\DW}^{r_\DW+d_\sph} dr \frac{d}{dr} \theta(r) = \epsilon\Delta\theta$,
and the total baryon number created in the lifetime of the black hole:
\begin{eqnarray}
 B &=& \int^{\tau_\BH} \: dt \: \dot{B}
   \;=\; - \frac{15}{4\pi^3 g_*} \: {\cal N} \kappa \alpha_\weak^5 \:
       \epsilon\Delta\theta \:
       \frac{m_\planck^2}{T_\weak T_\BH}.
\end{eqnarray}
This result has no parameters like $\beta$ and $T_\univ$.
Then we see the form of this result has a kind of stability.


We assume a scenario of the following three steps:
First, in the very early universe,
most of the matter existed as the primordial black holes
with its mass $m_\BH$,
i.e., the universe was black-hole-dominant.
Second,
the black holes evaporated through creating baryons in our processes.
Finally, after evaporation of these black holes,
the universe became radiation dominant at the temperature $T_\univ$
by the Hawking radiation from these black holes.
In this scenario,
we assumed the monochrome mass-spectrum of black holes
only for simplicity of calculation.
The scenario implies
age of the black-hole-dominant-universe $t_\univ$,
when the baryon number was created,
equals to lifetime of the black holes: $t_\univ \simeq \tau_\BH$.
Further more, it implies
the energy density of the black holes in the universe $\rho_\BH$
is transfered to
the energy density of radiation in the universe
$\rho_\radiation = \frac{\pi^2}{30} g_{*\univ} T_\univ^4$
by the Hawking radiation,
i.e., 
$\rho_\radiation \simeq \rho_\BH$ at the time $t_\univ$.
The Einstein equation of the black-hole-dominant-universe
is same as the one of the matter-dominant-universe
\cite{Barrow}\cite{Majumdar}\cite{KolbTurner}.
Then we have the relation
between the energy density and the age of the universe:
$\rho_\BH = \rho_{\rm matter dom.}
\equiv \frac{1}{6\pi} \frac{m_\planck^2}{t_\univ^2}$.
By combining the relations in the scenario and the expression of $\tau_\BH$,
we have a relation between
temperature of the universe and one of the black holes:
\begin{eqnarray}
 g_{*\univ}T_\univ^4
  &\simeq& \frac{\pi g_*^2}{80} \frac{T_\BH^6}{m_\planck^2},
  \label{eqTunivTbh}
\end{eqnarray}
and the energy density of the black holes:
$\rho_\BH \simeq \frac{\pi^3 g_*^2}{2400} \frac{T_\BH^6}{m_\planck^2}$.

In our theory,
the evaporation of the black holes in the Higgs phase vacuum
is essential to creating baryons,
and to conserve the created baryons,
the temperature of universe must be
lower than the electroweak critical temperature $T_\weak$.
By the relation (\ref{eqTunivTbh}),
this restriction gives us
an upper bound for the black-hole temperature
$T_\BH \lnear 4.0 \times 10^7 \GeV$
(see \fig{BHTemp.eps}).
Here we require that our theory does not affect
the very successful BBN theory \cite{KolbTurner}
at $T_\univ \simeq 10 \MeV \sim 0.1 \MeV$,
then we have a lower bound
for the black-hole temperature $T_\BH \gnear 5.9 \times 10^4 \GeV$
(see \fig{BHTemp.eps}).


When the black holes evaporate,
the number density of black holes is
$n_\BH = \rho_\BH / m_\BH
= \frac{\pi^4}{300} g_*^2 \frac{T_\BH^7}{m_\planck^4}$,
and the created baryon number density is $b = B \: n_\BH$.
At the same time,
the entropy density in the universe is
$s = \frac{2\pi^2}{45} g_{*\univ} T_\univ^3$.
By using relation (\ref{eqTunivTbh}),
we have the baryon-entropy ratios in the universe:
\begin{eqnarray}
 \frac{b}{s}
  &\simeq&
  \frac{45}{2 \pi^2 g_*} \;
  {\cal N} \kappa\alpha_\weak^5
  \epsilon \Delta\theta \:
  \frac{T_\univ}{T_\weak}. \nonumber\\
 &\simeq&
  7 \times 10^{-10} \times \frac{\Delta\theta}{\pi} \: \frac{T_\univ}{T_\weak},
\end{eqnarray}
when $3.9 \times 10^7 \GeV \gnear T_\BH \gnear 5.9 \times10^4 \GeV$.
We display this result in \fig{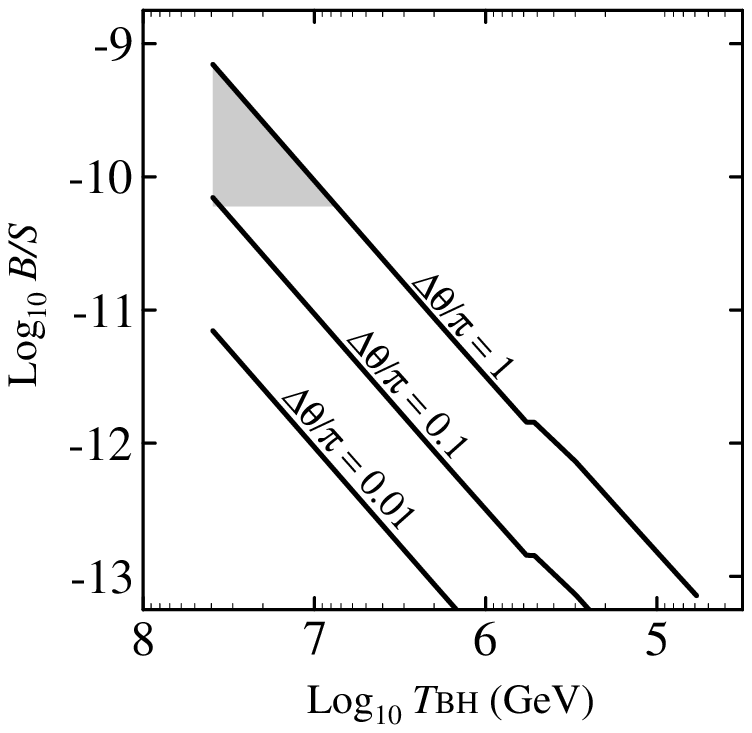}.

\begin{figure}
\begin{center}%
 \ \epsfbox{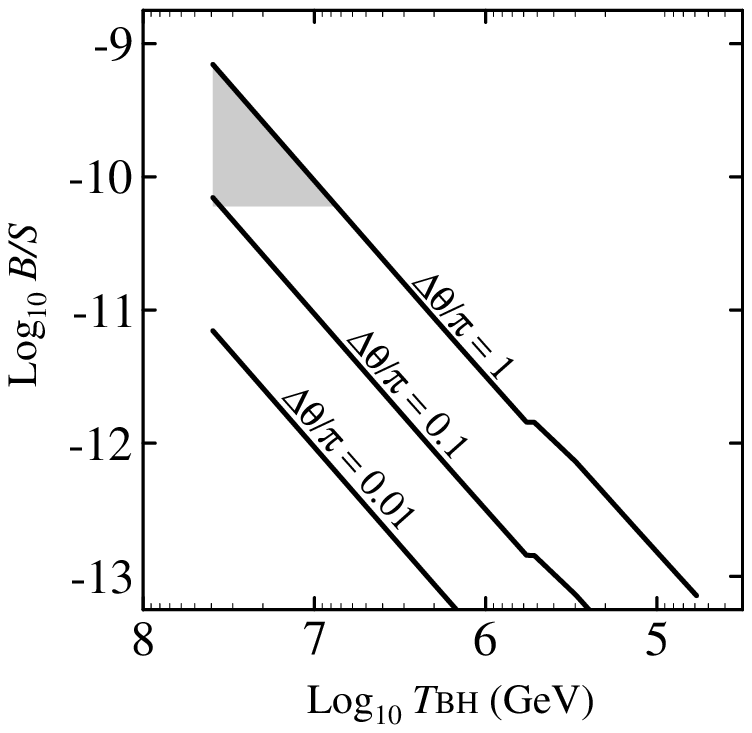}
 \caption{The resultant baryon-entropy ratio by our scenario
 	  when $\Delta\theta/\pi=1,0.1,0.01$.
 	  The shaded region satisfies the BBN requirement.
          }%
          \label{Result.eps}%
\end{center}%
\end{figure}

Finally,
if the domain wall has CP-phase: $\Delta\theta = \pi$,
as the most optimistic estimation,
the baryon-entropy ratio in our scenario
satisfies the BBN requirement: $B/S \sim 10^{-10}$ \cite{KolbTurner}
when $3.9 \times 10^7 \GeV \gnear T_\BH \gnear 7.4 \times 10^6 \GeV$,
namely $270\kg \lnear m_\BH \lnear 1400\kg$
(see the BBN-allowed region in \fig{Result.eps}).
In this parameter region,
we can neglect the diffusion enhancement effect \cite{CKN2},
because the width of domain wall is far greater than the electroweak scale.

In conclusion,
we have proposed a new scenario of the baryogenesis
which does not need the first order phase transition,
but does require the primordial black holes.


I would like to thank K. Yamawaki and A. I. Sanda
for helpful suggestions and discussions,
and also for careful reading the manuscript.
I also appreciate helpful suggestions of
A. Nakayama, M. Harada, T. Hanawa, K. Saigo and K. Shigetomi.
I am grateful to V. A. Miransky
for his encouragement and helpful advice. 
The work is supported in part
by a Grant-in-Aid for Scientific Research
from the Ministry of Education, Science and Culture
(No. 80003276).


\end{document}